\documentclass[reprint,amsmath,amssymb,aps]{revtex4-2}
\usepackage{graphicx}
\usepackage{dcolumn}
\usepackage{bm}
\usepackage{cleveref}
\usepackage{physics}
\def\iimg{{\bf i}}
\usepackage{float}
\usepackage{subfigure}
\usepackage{mathrsfs}

\begin{document}

\title{\bf{ Renormalization Group Approach for Modified vdP Oscillator with $\mathcal{PT}$ Symmetric Non-Hermitian Interaction}}
\author{Biswajit Bhowmick}
\email{biswajitbhowmick10@gmail.com, biswajitb22@iiserb.ac.in}
\affiliation{Department of Physics, Indian Institute for Science Education and Research Bhopal, Bhopal, India-462066}
\affiliation{Department of Physics, Institute of Science, Banaras Hindu University, Varanasi, India-221005}
\author{Rohit Mahendra Shinde}
\email{rohitshinde162@gmail.com, rohit\_ms@ph.iitr.ac.in}
\affiliation{Department of Physics, Indian Institute of Technology Roorkee, Roorkee, India-247667}
\affiliation{Department of Physics, Institute of Science, Banaras Hindu University, Varanasi, India-221005}
\author{Bhabani Prasad Mandal}
\email{bhabani.mandal@gmail.com}
\affiliation{Department of Physics, Institute of Science, Banaras Hindu University, Varanasi, India-221005}

\begin{abstract}
We consider a modified version of the well-known 2d vdP oscillator with a new non-Hermitian interaction. The usual perturbative approach fails to provide the classical dynamics of the system as the classical solutions become divergent in the long time limit. These kinds of divergences are similar to what occurs in quantum field theory and critical phenomena. The Renormalization Group (RG) technique for the dynamical system has been used to eliminate the divergences in the perturbative solution of the 2d vdP oscillator and to provide a physically acceptable solution which is shown to be consistent with numerical study. We further investigate the model in the framework of non-Hermitian quantum mechanics to show the $\mathcal{PT}$ phase transition in the system.
\end{abstract}

\maketitle

\section{Introduction}\label{intro}

\par The 1d van der Pol (vdP) oscillator was first introduced by Balthasar van der Pol, in the context of non-linear electrical circuits using vacuum tubes, specifically in triode circuits \cite{van1926lxxxviii}. It is described by a second-order ordinary differential equation with non-linear damping $\left(\mu\right)$ associated as follows,
\begin{align}
    \dv[2]{x}{t} - \mu \left(1-x^2\right)\dv{x}{t} + x = 0
\end{align}
The vdP equation has since been used to simulate diverse physical systems exhibiting periodic motions. This includes applications in non-linear electronic circuits \cite{robinson1987modified}, seismological applications to understand the earthquake faults \cite{cartwright1999dynamics}, biological applications to model electric potential across neuron cell membrane \cite{guckenheimer2000numerical}, etc. 

In this paper, we would like to study a $\mathcal{PT}$ symmetric non-Hermitain 2d vdP oscillator analytically as well as numerically in both classical and quantum mechanical frameworks. Over the past two and half decades, there has been great interest in a certain class of non-Hermitian quantum theories where the Hermiticity condition on the Hamiltonian of the system is relaxed with a physical but less constraining condition of Parity ($\mathcal{P}$) and Time reversal ($\mathcal{T}$) symmetry \cite{bender1998real, bender2007making, mostafazadeh2010pseudo, khare2000pt}. It has been shown that a consistent quantum theory with a complete real spectrum, unitary time evolution and probabilistic interpretation for $\mathcal{PT}$-symmetric non-Hermitian systems can be developed in a modified Hilbert space equipped with a positive-definite $\mathcal{CPT}$ inner product \cite{bender2002complex, das2010alternative}. $\mathcal{C}$ is an additional symmetry associated with every $\mathcal{PT}$-symmetric non-Hermitian system. Later $\mathcal{CPT}$ inner product is replaced by a more handy $\mathcal{G}$-inner product which is based on the geometry of quantum state \cite{mostafazadeh2007time, ju2019non, gardas2016non, tzeng2021hunting}. Because of this exciting realization, the research in non-Hermitian systems has received a huge boost over the past years \cite{Bender_1999,  BENDER_PTSymmetricQM, ghatak2013various, modak2021eigenstate, hasan2020role, shukla2023uncertainty,khare2009new,yadav2016parametric,basu2005phase, ohlsson2020transition, klauck2019observation, ashida2017parity, zhang2019time, bagarello2018bi, bagarello2021chain}. $\mathcal{PT}$-symmetric non-Hermitian systems have found numerous applications in various branches of physics and interdisciplinary areas \cite{pal2022dna, klaiman2008visualization, mandal2015pt, xu2016topological, xiao2016hyperbolic, musslimani2008optical, guo2009observation, west2010p, shukla2023uncertainty, khare2009new, yadav2016parametric, basu2005phase}, and some of the predictions of non-Hermitian theories are experimentally verified \cite{pan2018photonic, ozdemir2019parity, fleury2015invisible, musslimani2008optical,guo2009observation}.

Such $\mathcal{PT}$-invariant non-Hermitian systems generally exhibit a $\mathcal{PT}$ symmetry-breaking transition \cite{raval2018deconfinement, mandal2013pt, khare2000pt, mandal2015pt, pal2022dna} that separates two regions: (i) $\mathcal{PT}$-symmetric phase in which the entire spectrum is real and the eigenfunctions of the Hamiltonian respect $\mathcal{PT}$ symmetry and (ii) $\mathcal{PT}$-broken phase in which the entire spectrum (or a part of it) is in complex conjugate pairs and the eigenstates of the Hamiltonian are not the eigenstates of the $\mathcal{PT}$ operator. The phase transition happens at the exceptional point ($\mathcal{EP}$) for the particular Hamiltonian. While the $\mathcal{CPT}$ inner product is restricted to the $\mathcal{PT}$-symmetric phase only, the $\mathcal{G}$-metric inner product \cite{mostafazadeh2007time, ju2019non, gardas2016non, tzeng2021hunting} is extended even in the $\mathcal{PT}$-broken phase.

We consider the following version of the well-known 2d vdP oscillator with a non-Hermitian interaction,
\begin{align}
     \mathcal{H} &= p_xp_y + \omega^2xy - \iimg\left[\mu_1\left(1-x^2\right)yp_y+\mu_2\left(1-y^2\right)xp_x\right] \label{Hamiltonian}
\end{align}

The last term in the above Hamiltonian is non-Hermitian. Two real parameters $\mu_1$ and $\mu_2$  will play a crucial role in investigating the model in different phases of the system.
 We will see in the section \ref{MvdP-section} that in general \cref{Hamiltonian} remains non-$\mathcal{PT}$ symmetric but when $\mu_1=\mu_2$, $\mathcal{PT}$ symmetry is restored. In order to understand the dynamics of our considered Hamiltonian system, we take the perturbative approach to find its solutions. But inevitably, we see that perturbative calculations give rise to physically unaccepted solutions. The perturbative solutions diverge in the longer time limit. This kind of divergence arises in various other places in physics as in quantum field theory and in critical phenomenon \cite{wilson1983renormalization, pelissetto2002critical, weinberg1973new, zinn2021quantum}. To understand these divergences and to get physically meaningful solutions out of them, we take help from the renormalization group technique \cite{delamotte2004hint, 10.1119/1.4944701}. We explicitly demonstrate how the RG technique for dynamical systems leads to physical solutions consisting of limit cycles  \cite{strogatz2018nonlinear, he2005limit} and centres  \cite{strogatz2018nonlinear, Sarkar_2011, Sarkar_2012} depending on the conditions on the Hermiticity breaking parameters. Further, we obtain the numerical solutions of the 2d non-Hermitian vdP oscillator and show that these are completely consistent with the RG solutions of the system.  
In the quantum framework, we find the energy eigenvalues of the system numerically. The energy eigenvalues are complex in general and become real when the coupling is small in the $\mathcal{PT}$ unbroken phase. $\mathcal{PT}$ phase transition occurs when the Hermiticity breaking parameter exceeds a critical value.

Now we present the plan of the paper.  We first review the RG technique used in the perturbative study of the dynamical systems with the help of a simple model in section \ref{RG_Section}. In section \ref{MvdP-section}, we consider the solution of the 2d vdP oscillator by both perturbative as well as RG method. The equivalence of RG solution with the numerical solution of 2d vdP oscillator is established in section \ref{Comparison}. Next, we demonstrate the quantum mechanical aspects of our system along with the consequences of $\mathcal{PT}$ phase transition in section \ref{QM_MvdP}. The last section is dedicated to summary and discussion.

\section{Review of the Renormalization Group Techniques}\label{RG_Section}

Following Delamotte \cite{delamotte2004hint} in this section, we briefly review the essential mathematical details of RG techniques for the dynamical system with a simple example.  We consider a Hamiltonian system consisting of a single non-quadratic term (interaction term) characterized by some dimensionless coupling constant $g_0$. Let us now define a physical observable $F(x)$ which we would like to calculate perturbatively in terms of the coupling constant $g_0$. We assume a power series expansion of $F(x)$ in terms of $g_0$ as,

\begin{align}
    F(x) &= g_0 + g_0^2F_1(x) + \mathcal{O}\left(g_0^3\right) \label{F_pert}
\end{align}

Let us further assume that the coefficient of $g_0^2$ in \cref{F_pert} has a logarithmic divergence of the following form, 
\begin{align}
    F_1(x) &= \int_{0}^{\infty} \frac{dt}{t+x}\label{F_1}
\end{align}
as typically arises in quantum field theory.

 However physical observables, $F(x) $ cannot be divergent in nature and the only thing that might have gone wrong is the choice of the coupling constant $g_0$, everything else is restricted by the dynamics and symmetry of the given system.  To make sense of the theory one has to take help from experimental results and fix the value of $g_0$ such that for say some $x=\mu$, $F(\mu)$ given by the theory matches with the finite result of the experiment.  It becomes evident that the primary issue does not reside within $F_i(x)$ or the perturbative expansion; rather, it lies in the choice of $g_0$. Therefore we define the new coupling constant by using the finite experimental result of $F(\mu)$,

\begin{align}
    F(\mu) &= g_r \label{F_mu}
\end{align}

We introduce a renormalization constant $Z$ which connects this new coupling constant to $g_0$ as,

\begin{align}
    g_0 &= Z g_r\label{g0_gr}
\end{align}

 and define a power series expansion of $Z$ in terms of $g_r$ as,

\begin{align}
    Z &= 1 + z_1 g_r + z_2 g_r^2 + \mathcal{O}\left(g_r^3\right)\label{Z}.
\end{align}

To remove the infinities in \cref{F_1} we  write the regularized form of $F(x)$ as,

\begin{align}
    F_{\Lambda}\left(x, g_0,\Lambda\right) &= g_0 + g_0^2 \ln{\left(\frac{\Lambda+x}{x}\right)}+ \mathcal{O}\left(g_0^3\right) 
\end{align}

where we have used the cut off ($\Lambda$) regularization of $F_1(x)$ given in  \cref{F_1}. Further, this can be expressed in terms of renormalized coupling using \cref{g0_gr,Z} as,

\begin{align}
    F_{\Lambda}\left(x,g_r,\Lambda\right) &= g_r + g_r^2 \left[ z_1 + \ln{\left(\frac{\Lambda+x}{x}\right)} \right] + \mathcal{O}\left(g_r^3\right)\label{F_Lambda}
\end{align}

The \cref{F_mu} will be assured at $x=\mu$ by choosing

\begin{align}
    z_1 &= -\ln{\left(\frac{\Lambda+\mu}{\mu}\right)}\label{z_1} 
\end{align}
 Using \cref{z_1} in \cref{F_Lambda}, we have

\begin{align}
    F_{\Lambda}\left(x,g_r,\Lambda\right) &= g_r + g_r^2 \left[ -\ln{\left(\frac{\Lambda+\mu}{\mu}\right)} + \ln{\left(\frac{\Lambda+x}{x}\right)} \right]\nonumber \\
	&\mathrel{\phantom{=}} + \mathcal{O}\left(g_r^3\right)
\end{align}

Taking the limit, $\Lambda\to\infty$ we find that the above expression becomes,

\begin{align}
\lim_{\Lambda\to\infty}F_{\Lambda}\left(x,g_r,\Lambda\right) &= F\left(x,g_r,\mu\right)\\ \nonumber
 &= g_r - g_r^2\ln{\left(\frac{x}{\mu}\right)} + \mathcal{O}\left(g_r^3\right)
\end{align}

As per our claim $F\left(x,g_r,\mu\right)$ is finite and independent of choice of the scale $\mu$, leading to the RG flow equation
\begin{align}
    \pdv{F\left(x,g_r,\mu\right) }{\mu} =0 
\end{align}

This  technique can be illustrated with a simple   differential equation of $y(t)$, as given below,

\begin{align}
    \dv{y}{t}+\epsilon y= 0 \label{eg_DE_y}
\end{align}
where $\epsilon$ is a small constant parameter of the system. Assuming the initial conditions for this system as $y\left(t=0\right)=A$, we can see the exact solution becomes $y(t)=Ae^{-\epsilon t}$. Let's try to solve this differential equation using the perturbative method by assuming a power series expansion of $y(t)$ in terms of the parameter $\epsilon$, such that

\begin{align}
    y &= y_0 + \epsilon y_1 + \mathcal{O}\left( \epsilon^2\right)\label{eg_y_series}
\end{align}

Then the DE in \cref{eg_DE_y} can equivalently be described  (considering only up to $\mathcal{O}(\epsilon)$) by,

\begin{align}
    \dv{y_0}{t} &=0 \label{eg_DE_y0}\\
    \dv{y_1}{t} + y_0 &=0 \label{eg_DE_y1}
\end{align}

 The initial conditions for $y(t)$ for some non-zero time $t=t_0$, $y(t_0)=A\left(t_0\right)$ can be written as $y_0(t_0)=A\left(t_0\right)$ and $y_n(t_0)=0$ $\forall \, n>0$. With these initial conditions the DEs in  \cref{eg_DE_y0,eg_DE_y1} admit the solutions,

\begin{align}
    y_0(t) &= A\left(t_0\right)\\
    y_1(t) &=  -A\left(t_0\right)\left( t-t_0 \right)
\end{align}

and thus the solution for $y(t)$ is,

\begin{align}
    y(t) &= A\left( t_0 \right) -\epsilon A\left(t_0\right)\left( t-t_0 \right) + \mathcal{O}\left(\epsilon^2\right)\label{eg_pert_ysol}
\end{align}
The perturbative method to find a solution for \cref{eg_DE_y} fails as
the solution diverges when $\left(t-t_0\right)\to\infty$. The only thing that might have gone wrong in this analysis is the choice of the initial condition. So instead of choosing our initial condition at $t=t_0$ we equivalently can choose it at some other point $t=\tau$ in the trajectory, such that we have $y\left(t=\tau\right)=A\left(\tau\right)$ which again translates to $y_0(\tau)=A\left(\tau\right)$ and $y_n(\tau)=0$ $\forall \, n>0$. We now define a relationship between the initial conditions at $t=t_0$ and that at $t=\tau$ using a renormalization constant $\mathcal{Z}\left(t_0,\tau\right)$ as,

\begin{align}
    A(t_0) &= \mathcal{Z}\left(t_0,\tau\right) A(\tau)\label{eg_At0_Atau}
\end{align}

We consider a power series expansion of $\mathcal{Z}\left(t_0,\tau\right)$ in terms of $\epsilon$ as,

\begin{align}
    \mathcal{Z}\left(t_0,\tau\right) &= 1 + \epsilon a_1 + \mathcal{O}
    \left(\epsilon^2\right) \label{eg_Z}
\end{align}

The idea is to redefine $a_1$ in the above equation to remove the divergences from our perturbative solution to get a physical solution. Substituting \cref{eg_Z,eg_At0_Atau} in  \cref{eg_pert_ysol} we get,
\begin{align}
    y(t) &= A\left(\tau\right) - \epsilon \left(t-\tau\right) A\left(\tau\right) + \mathcal{O}\left(\epsilon^2\right)\label{eg_ysol_RG_betn2}
\end{align}
where $a_1$ is chosen as,

\begin{align}
    a_1 &= \tau - t_0
\end{align}

The physical solution cannot depend on some arbitrarily chosen initial conditions at $t=\tau$, this leads to the flow equation,

\begin{align}
    \dv{y}{\tau} &=0\label{eg_FlowEq}
\end{align}

 Now using \cref{eg_ysol_RG_betn2,eg_FlowEq} we find that,

\begin{align}
    \dv{A\left(\tau\right)}{\tau} + \epsilon A\left(\tau\right) &= 0
\end{align}

which has the following solution with the initial condition that at $\tau=0$, $A(\tau)=A=\text{constant}$,

\begin{align}
    A\left(\tau\right) &= A e^{-\epsilon\tau}
\end{align}

With this \cref{eg_ysol_RG_betn2} takes the following form,

\begin{align}
    y\left(t\right) &= A e^{-\epsilon \tau} \left[ 1 -\epsilon\left( t- \tau\right)\right] + \mathcal{O}\left(\epsilon^2\right)
\end{align}

Now as $\tau$ could be chosen arbitrarily, the best choice is to consider $\tau=t$ such that our solution $y(t)$ takes the form,

\begin{align}
    y\left(t\right) &= A e^{-\epsilon t}
\end{align}

This demonstrates how to obtain physical solutions using the RG technique. We will be applying this technique to our 2d vdP oscillator problem in the next section.

\section{Modified 2d vdP oscillator}\label{MvdP-section}

In this section, we consider the modified vdP oscillator with the non-Hermitian interaction described by
\cref{Hamiltonian}. This Hamiltonian is invariant under the following $\mathcal{PT}$ transformations  in 2d, when $\mu_1=\mu_2 $,
\begin{align*}
\mathcal{P}: && x\longrightarrow y ; \  y\longrightarrow x ;\ 
p_x\longrightarrow p_y ; \  p_y\longrightarrow p_x \nonumber \\
\mathcal{T}: && i\longrightarrow -i ; \   
p_x\longrightarrow -p_x ; \  p_y\longrightarrow - p_y 
\end{align*}
First, we would like to investigate the system classically to explore any notable differences in the solutions of $\mathcal{PT}$ symmetric $(\mu_1=\mu_2)$ and non-$\mathcal{PT}$ symmetric $(\mu_1\ne \mu_2)$ situations. Keeping this in mind we write down the classical equations of motion as,

\begin{align}
	\ddot{x}&=-\omega^2x+\iimg\mu_1(1-x^2)\dot{x}-\iimg\mu_2(1-y^2)\dot{x} \nonumber \\
	&\mathrel{\phantom{=}} -\mu_1\mu_2\left(x(1-x^2)(1-y^2)-2xy^2(1-x^2)\right) \label{xddot}
\end{align}

\begin{align}
	\ddot{y}&=-\omega^2y-\iimg\mu_1(1-x^2)\dot{y}+\iimg\mu_2(1-y^2)\dot{y} \nonumber \\
	&\mathrel{\phantom{=}} -\mu_1\mu_2\left(y(1-x^2)(1-y^2)-2x^2y(1-y^2)\right)\label{yddot}
\end{align}
These are coupled non-linear differential equations and can't be solved analytically. So, we use the perturbative approach to find the solutions for small $\mu_1$ and $\mu_2$.

\subsection{Perturbative Solution}\label{PertSoln}

We assume a perturbative expansion of $x$ and $y$ in terms of $\mu_1$ and $\mu_2$ of the following form,
\begin{equation}\label{expansion_pert}
    x= \sum_{i,j=0}^{\infty}\mu_1^i\mu_2^jx_{ij}, \ \ y= \sum_{i,j=0}^{\infty}\mu_1^i\mu_2^jy_{ij} 
    \end{equation}
    Keeping to the lowest order we can write,
\begin{align}
	x&=x_{00}+\frac{\mu_1}{2}x_{10}+\frac{\mu_2}{2}x_{01}+\mathcal{O}(\mu^2)\label{perx}\\
	y&=y_{00}+\frac{\mu_1}{2}y_{10}+\frac{\mu_2}{2}y_{01}+\mathcal{O}(\mu^2)\label{pery}
\end{align}

where $\mathcal{O}(\mu^2)$ refers to powers of $\mu_1^m\mu_2^n$ with $m+n\ge 2$. Now, substituting \cref{perx,pery} back in the classical equations of motion, that is in \cref{xddot,yddot} then equating terms with the same powers of $\mu_1^m\mu_2^n$ where $(m,n=0,1)$, we get at different orders, for $x_{mn}(t)$ and $y_{mn}(t)$ the following,

\begin{align}
	\mu^0_1;\mu^0_2: \qquad \ddot{x}_{00}&=-\omega^2x_{00}\label{ddotx00}\\
	\mu^1_1;\mu^0_2: \qquad \ddot{x}_{10}&=-\omega^2x_{10} +2\iimg(1-x^2_{00})\dot{x}_{00}\label{ddotx10}\\
	\mu^0_1;\mu^1_2: \qquad \ddot{x}_{01}&=-\omega^2x_{01} -2\iimg(1-y^2_{00})\dot{x}_{00}\label{ddotx01}
\end{align}

\begin{align}
	\mu_1^0;\mu_2^0 :\qquad \ddot{y}_{00}&=-\omega^2\ddot{y}_{00}\label{y_00}\\
	\mu_1^1;\mu_2^0 :\qquad  \ddot{y}_{10}&=-\omega^2y_{10}-2\iimg(1-x_{00}^2)\dot{y}_{00}\label{y_10}\\
	\mu_1^0;\mu_2^1 :\qquad  \ddot{y}_{01}&=-\omega^2y_{01}+2\iimg(1-y_{00}^2)\dot{y}_{00}\label{y_01}
\end{align}

Let us consider the initial condition  specified at some  $t=t_0$ as $x(t_0)=A(t_0)$ and $\dot{x}(t_0)=0$, and for $y$, $y(t_0)=B(t_0)$ and $\dot{y}(t_0)=0$, where $A\left(t_0\right)$ and $B\left(t_0\right)$ are some constants. These initial conditions for $x$  and $y$ can be rewritten as, $x_{00}(t_0)=A(t_0),\ x_{mn}(t_0)=0$ for all $m,n \ge 1$, $\dot{x}_{m,n}(t_0)=0$ and $y_{00}(t_0)=B(t_0), y_{m,n}(t_0) =0 $ for all $m,n \ge 1$, $\dot{y}_{m,n}(t_0)=0$ respectively  for the components in the perturbative expansion. The set of coupled nonlinear differential equations given before are solved using the above-stated initial conditions (details of that are provided in Appendix \ref{Appendix}). The solutions for $x$ and $y$ up to the first order in $\mu_1$ and $\mu_2$ are given as,

\begin{align}
	x(t)&=A(t_0)\cos\left(\omega(t-t_0)\right)\nonumber \\
	&\mathrel{\phantom{=}}-\frac{\iimg\mu_1A(t_0)}{8}\left[(-4+A^2(t_0))(t-t_0)\right]\cos\left(\omega(t-t_0)\right) \nonumber \\
	&\mathrel{\phantom{=}} -\frac{\iimg\mu_1A(t_0)}{16\omega}\left[8-3A^2(t_0)+A^2(t_0)\right]\nonumber \\
	&\mathrel{\phantom{=}}\times\cos\left(2\omega(t-t_0)\right)\sin\left(\omega(t-t_0)\right) \nonumber \\ 
	&\mathrel{\phantom{=}} +\frac{\iimg\mu_2A(t_0)}{8}\left[(-4+B^2(t_0))(t-t_0)\right]\cos\left(\omega(t-t_0)\right) \nonumber \\
	&\mathrel{\phantom{=}} +\frac{\iimg\mu_2A(t_0)}{16\omega}\left[8-3B^2(t_0)+B^2(t_0)\right]\nonumber \\
	&\mathrel{\phantom{=}}\times\cos\left(2\omega(t-t_0)\right)\sin\left(\omega(t-t_0)\right) + \mathcal{O}\left(\mu^2\right) \label{x_sol}
\end{align}

\begin{align}
	y(t)&=B(t_0)\cos\left(\omega(t-t_0)\right)\nonumber \\
	&\mathrel{\phantom{=}}+\frac{\iimg\mu_1B(t_0)}{8}\left[(-4+A^2(t_0))(t-t_0)\right]\cos\left(\omega(t-t_0)\right) \nonumber \\
	&\mathrel{\phantom{=}} +\frac{\iimg\mu_1B(t_0)}{16\omega}\left[8-3A^2(t_0)+A^2(t_0)\right]\nonumber \\
	&\mathrel{\phantom{=}}\times\cos\left(2\omega(t-t_0)\right)\sin\left(\omega(t-t_0)\right) \nonumber \\ 
	&\mathrel{\phantom{=}} -\frac{\iimg\mu_2B(t_0)}{8}\left[(-4+B^2(t_0))(t-t_0)\right]\cos\left(\omega(t-t_0)\right) \nonumber \\
	&\mathrel{\phantom{=}} -\frac{\iimg\mu_2B(t_0)}{16\omega}\left[8-3B^2(t_0)+B^2(t_0)\right]\nonumber \\
	&\mathrel{\phantom{=}}\times\cos\left(2\omega(t-t_0)\right)\sin\left(\omega(t-t_0)\right) + \mathcal{O}\left(\mu^2\right) \label{y_sol}
\end{align}

\subsection{Renormalization Group aided solution}

We find that the solutions for $x\left(t\right)$ and $y\left(t\right)$ given by \cref{x_sol,y_sol} in the previous section are not well defined as some terms diverge 
 as $\left(t-t_0\right)\to\infty$. This is similar to what happened in \cref{eg_pert_ysol}. Thus the perturbation approach fails to give us physical solutions. So as before, the only thing that might have gone wrong in this analysis is our choice of initial conditions, everything else was predefined and restricted by our Hamiltonian and thus by the dynamics of our system. So instead of choosing our initial condition at $t=t_0$, we can equivalently choose them at some other time $t=\tau$, such that we will write our solutions in terms of $A(\tau)$, $B(\tau)$ and $\theta(\tau)$. Here we consider $\theta\left(\tau\right)$ as renormalized form of $-t_0$ which we again call as $\theta\left(t_0\right)$ for consistency. We now define the relationship between the initial conditions at $t=t_0$ and $t=\tau$ as like in \cref{eg_At0_Atau} by using renormalization constants $\mathcal{Z}_A $, $\mathcal{Z}_B$ and $\mathcal{Z}_\theta$ as,

\begin{align}
	A(t_0)&=A(\tau)\mathcal{Z}_A(t_0,\tau)\label{A_tau}\\
	B(t_0)&=B(\tau)\mathcal{Z}_B(t_0,\tau)\label{B_tau} \\
	-t_0=\theta(t_0)&=\theta(\tau)+\mathcal{Z}_\theta(t_0,\tau) \label{theta_tau}
\end{align}

Here we take the relations between the amplitudes as multiplicative as it occurs in our solution as a multiplicative factor, but for the case of time, we considered an additive relation to make it consistent with the fact that it started from some point. Now we consider a power series expansion of the renormalization constants of the form,

\begin{align}
\mathcal{Z}_A(t_0,\tau)&=1+\frac{\mu_1}{2}a_{10}+\frac{\mu_2}{2}a_{01}+\mathcal{O}(\mu^2)\label{Z_A}\\
\mathcal{Z}_B(t_0,\tau)&=1+\frac{\mu_1}{2}b_{10}+\frac{\mu_2}{2}b_{01}+\mathcal{O}(\mu^2)\label{Z_B}\\
\mathcal{Z}_\theta(t_0,\tau)&=\frac{\mu_1}{2}c_{10}+\frac{\mu_2}{2}c_{01}+\mathcal{O}(\mu^2)\label{Z_theta}
\end{align}

The goal is to use these constants $a_{ij}$, $b_{ij}$ and $c_{ij}$ in such a way that it removes the long-time divergences in our solutions in \cref{x_sol,y_sol}. We then proceed in a very similar fashion as was done in section \ref{RG_Section} and find that to remove the diverging terms of the form $\left(t-t_0\right)$, we have to choose $a_{10}$, $a_{01}$, $b_{10}$ and $b_{01}$ as,

\begin{align}
a_{10}&=\frac{\iimg}{4}\left[-4+A^2(\tau)\right](\tau-t_0) \label{a10} \\
a_{01}&=-\frac{\iimg}{4}\left[-4+B^2(\tau)\right](\tau-t_0)\label{a01}\\
b_{10}&=-\frac{\iimg}{4}\left[-4 + A^2(\tau)\right](\tau-t_0)\label{b_10}\\
b_{01}&=\frac{\iimg}{4}\left[-4+B^2(\tau)\right](\tau-t_0)\label{b_01}
\end{align}

 The solutions then  take the following form,

\begin{align}
x(t)&= A(\tau)\cos\left(\omega(t+\theta(\tau))\right) \nonumber \\
&\mathrel{\phantom{=}} -\frac{\iimg\mu_1A(\tau)}{8}\left[(-4+A^2(\tau))(t-\tau)\right]\cos\left(\omega(t+\theta(\tau))\right)\nonumber \\
&\mathrel{\phantom{=}} -\frac{\iimg\mu_1A(\tau)}{16\omega}\left[8-3A^2(\tau)+A^2(\tau)\right]\nonumber \\
	&\mathrel{\phantom{=}}\times\cos(2\omega(t+\theta(\tau)))\sin(\omega(t+\theta(\tau)))\nonumber \\
&\mathrel{\phantom{=}} +\frac{\iimg\mu_2A(\tau)}{8}\left[(-4+B^2(\tau))(t-\tau)\right]\cos(\omega(t+\theta(\tau)))\nonumber \\
&\mathrel{\phantom{=}} +\frac{\iimg\mu_2A(\tau)}{16\omega}\left[8-3B^2(\tau)+B^2(\tau)\right]\nonumber \\
	&\mathrel{\phantom{=}}\times\cos(2\omega(t+\theta(\tau)))\sin(\omega(t+\theta(\tau)))+ \mathcal{O}(\mu^2)\label{x_sol_RG}
\end{align}
\begin{align}
y(t)&=B(\tau)\cos(\omega(t+\theta(\tau)))\nonumber\\
&\mathrel{\phantom{=}}+\frac{\iimg\mu_1 B(\tau)}{8}\left[(-4+A^2(\tau))(t-\tau)\right]\cos(\omega(t+\theta(\tau)))\nonumber\\
&\mathrel{\phantom{=}} +\frac{\iimg\mu_1 B(\tau)}{16\omega}\left[8-3A^2(\tau)+A^2(\tau)\right]\nonumber \\
	&\mathrel{\phantom{=}}\times\cos(2\omega(t+\theta(\tau)))\sin(\omega(t+\theta(\tau))) \nonumber\\
&\mathrel{\phantom{=}} -\frac{\iimg\mu_2 B(\tau)}{8}\left[(-4+B^2(\tau))(t-\tau)\right]\cos(\omega(t+\theta(\tau)))\nonumber\\
&\mathrel{\phantom{=}} -\frac{\iimg\mu_2 B(\tau)}{16\omega}\left[(8-3B^2(\tau)+B^2(\tau))\right]\nonumber \\
	&\mathrel{\phantom{=}}\times\cos(2\omega(t+\theta(\tau)))\sin(\omega(t+\theta(\tau)))+ \mathcal{O}(\mu^2)\label{y_sol2}
\end{align}

\subsection{Flow Equation}

The physical solutions for $x(t)$ and $y(t)$ as given in \cref{x_sol_RG,y_sol2} cannot depend on some arbitrarily chosen initial conditions at $t=\tau$ and hence we require that,

\begin{align}
	\dv{x}{\tau}=0 ,\qquad	\dv{y}{\tau}=0 \label{deri_x_0}
\end{align}

The above flow equations can be realized in two different ways as shown in the Appendix \ref{Appendix}, namely  (i) When  $\dv{\theta(\tau)}{\tau}=0 $  and when
(ii) $B(\tau)=\pm\sqrt{\frac{\mu_1}{\mu_2}}A(\tau)$. For the former case, the physical solutions are given by,

\begin{align}
    x(t)&=A(t_0)\cos(\omega(t-t_0)) \nonumber \\
&\mathrel{\phantom{=}} -\frac{\iimg\mu_1A(t_0)}{16\omega}\left[8-3A^2(t_0)+A^2(t_0)\right]\nonumber \\
	&\mathrel{\phantom{=}}\times\cos\left(2\omega(t-t_0)\right)\sin\left(\omega(t-t_0)\right)\nonumber \\
&\mathrel{\phantom{=}} +\frac{\iimg\mu_2A(t_0)}{16\omega}\left[8-3B^2(t_0)+B^2(t_0)\right]\nonumber \\
	&\mathrel{\phantom{=}}\times\cos\left(2\omega(t-t_0)\right)\sin\left(\omega(t-t_0)\right)\nonumber\\
&\mathrel{\phantom{=}}+ \mathcal{O}(\mu^2)\label{x_sol_RGfinal_1}
\end{align}

\begin{align}
    y(t)&=B(t_0)\cos(\omega(t-t_0))\nonumber\\
&\mathrel{\phantom{=}}+\frac{\iimg\mu_1 B(t_0)}{16\omega}\left[8-3A^2(t_0)+A^2(t_0)\right]\nonumber \\
	&\mathrel{\phantom{=}}\times\cos\left(2\omega(t-t_0)\right)\sin\left(\omega(t-t_0)\right)\nonumber\\
&\mathrel{\phantom{=}} -\frac{\iimg\mu_2 B(t_0)}{16\omega}\left[8-3B^2(t_0)+B^2(t_0)\right]\nonumber \\
	&\mathrel{\phantom{=}}\times\cos\left(2\omega(t-t_0)\right)\sin\left(\omega(t-t_0)\right)\nonumber\\
&\mathrel{\phantom{=}}+ \mathcal{O}(\mu^2)\label{y_sol_RGfinal_1}
\end{align}

For the latter case, we have the following solutions,

\begin{align}
x(t)&=A(t_0)\cos\left(\alpha(t-t_0)\right)\cos\left(\beta\left(t-t_0\right)\right) \nonumber \\
&\mathrel{\phantom{=}} +\frac{\mu_1}{2\omega}A(t_0)\sin(\alpha(t-t_0))\sin(\beta\left(t-t_0\right)) \nonumber \\
&\mathrel{\phantom{=}}  -\frac{\mu_2}{2\omega}A(t_0)\sin(\alpha(t-t_0))\sin(\beta\left(t-t_0\right))\nonumber \\
&\mathrel {\phantom{=}} +\iimg\Big[A(t_0)\sin(\alpha(t-t_0))\cos(\beta\left(t-t_0\right)) \nonumber \\
&\mathrel{\phantom{=}} -\frac{\mu_1}{2\omega}A(t_0)\cos(\alpha(t-t_0))\sin(\beta\left(t-t_0\right))\nonumber \\
&\mathrel{\phantom{=}}  +\frac{\mu_2}{2\omega}A(t_0)\cos(\alpha(t-t_0))\sin(\beta\left(t-t_0\right)) \Big]\nonumber\\
&\mathrel{\phantom{=}}+ \mathcal{O}(\mu^2)\label{x_sol_RGfinal_2}
\end{align}

\begin{align}
y(t)&= \pm\sqrt{\frac{\mu_1}{\mu_2}}A(t_0)\cos(\alpha(t-t_0))\cos(\beta\left(t-t_0\right))\nonumber\\
&\mathrel{\phantom{=}} \pm\frac{\mu_1}{2\omega}\sqrt{\frac{\mu_1}{\mu_2}}A(t_0)\sin(\alpha(t-t_0))\sin(\beta\left(t-t_0\right))\nonumber\\
&\mathrel{\phantom{=}} \mp\frac{\mu_2}{2\omega}\sqrt{\frac{\mu_1}{\mu_2}}A(t_0)\sin(\alpha(t-t_0) )\sin(\beta\left(t-t_0\right))\nonumber\\
&\mathrel{\phantom{=}} +\iimg\left[ \mp\sqrt{\frac{\mu_1}{\mu_2}}A(t_0)\sin(\alpha(t-t_0))\cos(\beta\left(t-t_0\right))\right.\nonumber\\
&\mathrel{\phantom{=}} \mp\frac{\mu_1}{2\omega}\sqrt{\frac{\mu_1}{\mu_2}}A(t_0)\cos(\alpha(t-t_0))\sin(\beta\left(t-t_0\right))\nonumber\\
&\mathrel{\phantom{=}}\left. \pm\frac{\mu_2}{2\omega}\sqrt{\frac{\mu_1}{\mu_2}}A(t_0)\cos(\alpha(t-t_0))\sin(\beta\left(t-t_0\right))\right]\nonumber\\
&\mathrel{\phantom{=}}+ \mathcal{O}(\mu^2)\label{y_sol_RGfinal_2}
\end{align}

where we have denoted $\alpha$ and $\beta$  for brevity as,

\begin{align}
    \alpha &= \frac{ 2\omega^2(\mu_1-\mu_2)}{4\omega^2-(\mu_1-\mu_2)^2}\\
    \beta &= \frac{4\omega^3}{4\omega^2-(\mu_1-\mu_2)^2}
\end{align}
and have used the following initial conditions,

\begin{align}
    \theta(\tau=t_0)&=-t_0\\
    A(\tau=t_0)&=A(t_0)=\text{constant}\\
    B(\tau=t_0)&=B(t_0)=\text{constant}
\end{align}

Notice the RG-aided solutions in \cref{x_sol_RGfinal_1,y_sol_RGfinal_1} and \cref{x_sol_RGfinal_2,y_sol_RGfinal_2} do not have any divergent term of the form $\left(t-t_0\right)$, thus giving us the physical solutions for our dynamical system.  We will be considering the solutions given by \cref{x_sol_RGfinal_2,y_sol_RGfinal_2} for further analysis. 

\section{Comparison between RG-aided Solutions and Numerical Solutions}\label{Comparison}

\begin{figure}[h]
     \begin{center}
        \subfigure[]{%
            \label{fig:first1}
            \includegraphics[width=0.2\textwidth]{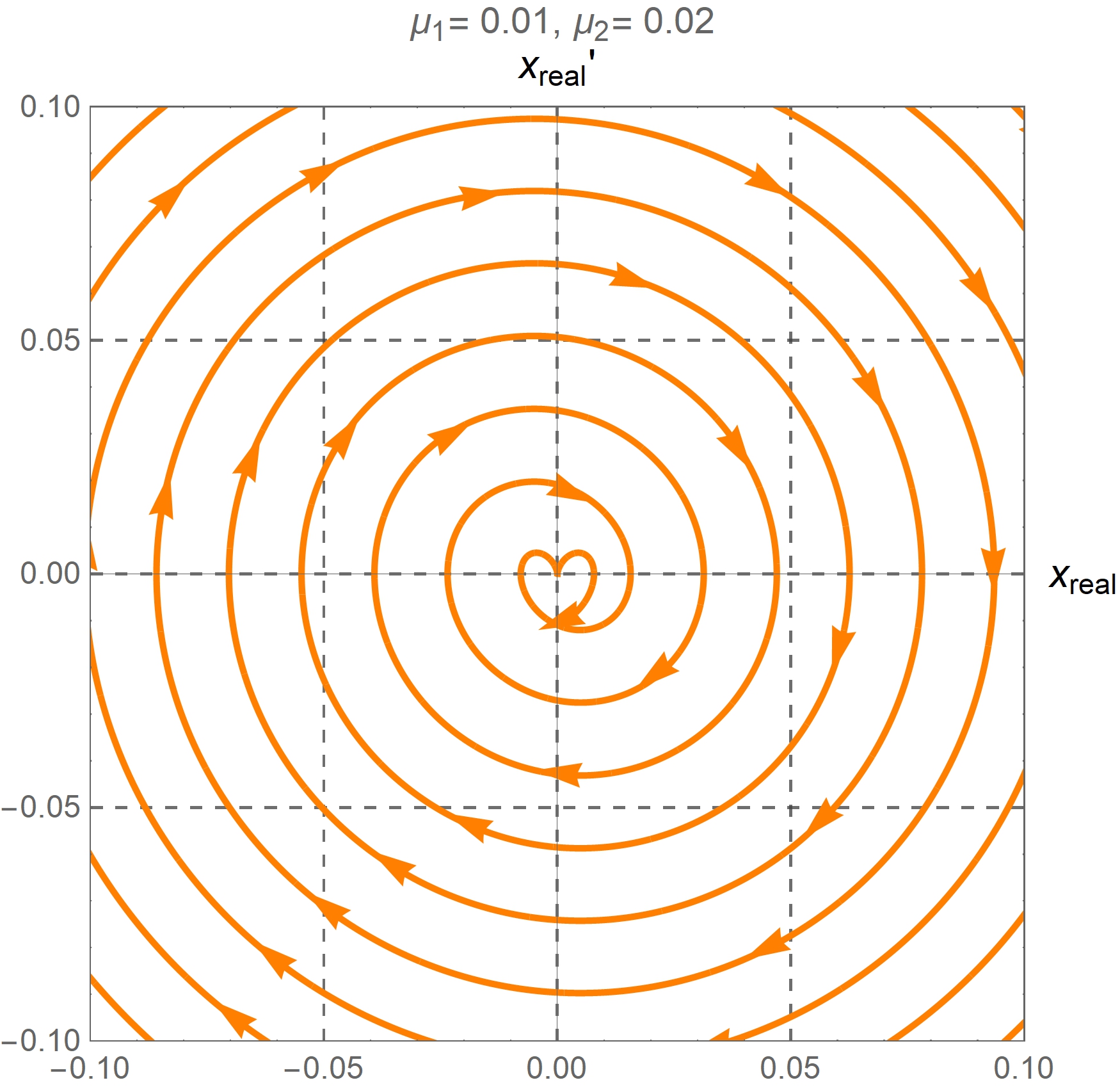}
        }%
        \subfigure[]{%
           \label{fig:second1}
           \includegraphics[width=0.2\textwidth]{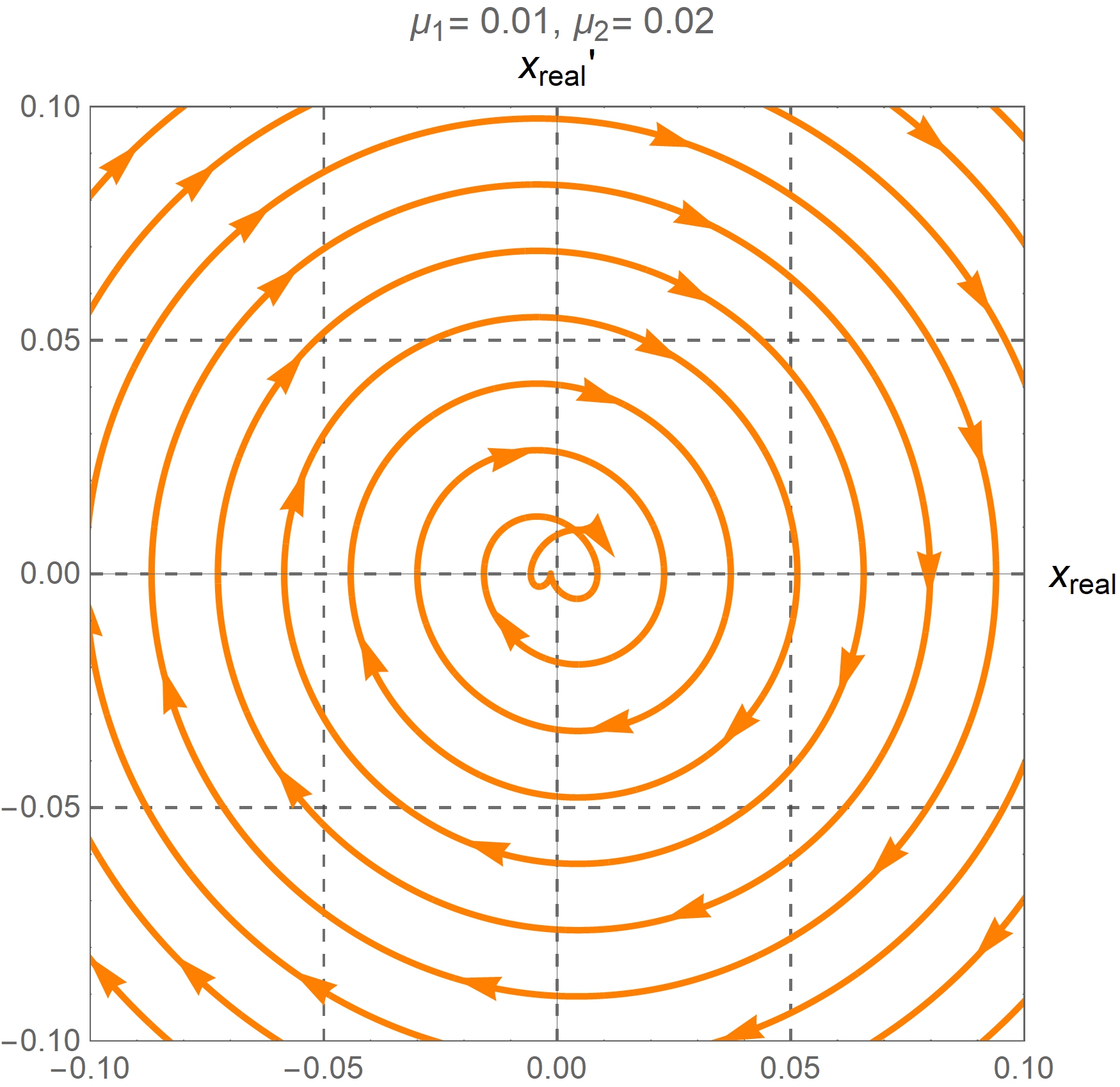}
        } \\
        
    \end{center}
    \caption{Time derivative of real part of $x(t)$ is plotted with real part of $x(t)$. (a) RG-aided Solution and (b) Numerical Solution. Here we have chosen $\mu_1=0.01$ and $\mu_2=0.02$ $\left(\mu_1\neq\mu_2\right)$ which is the non-$\mathcal{PT}$ symmetric case.}
   \label{fig:subfigures i1}
\end{figure}

\begin{figure}[h]
     \begin{center}
        \subfigure[]{%
            \label{fig:first5}
            \includegraphics[width=0.2\textwidth]{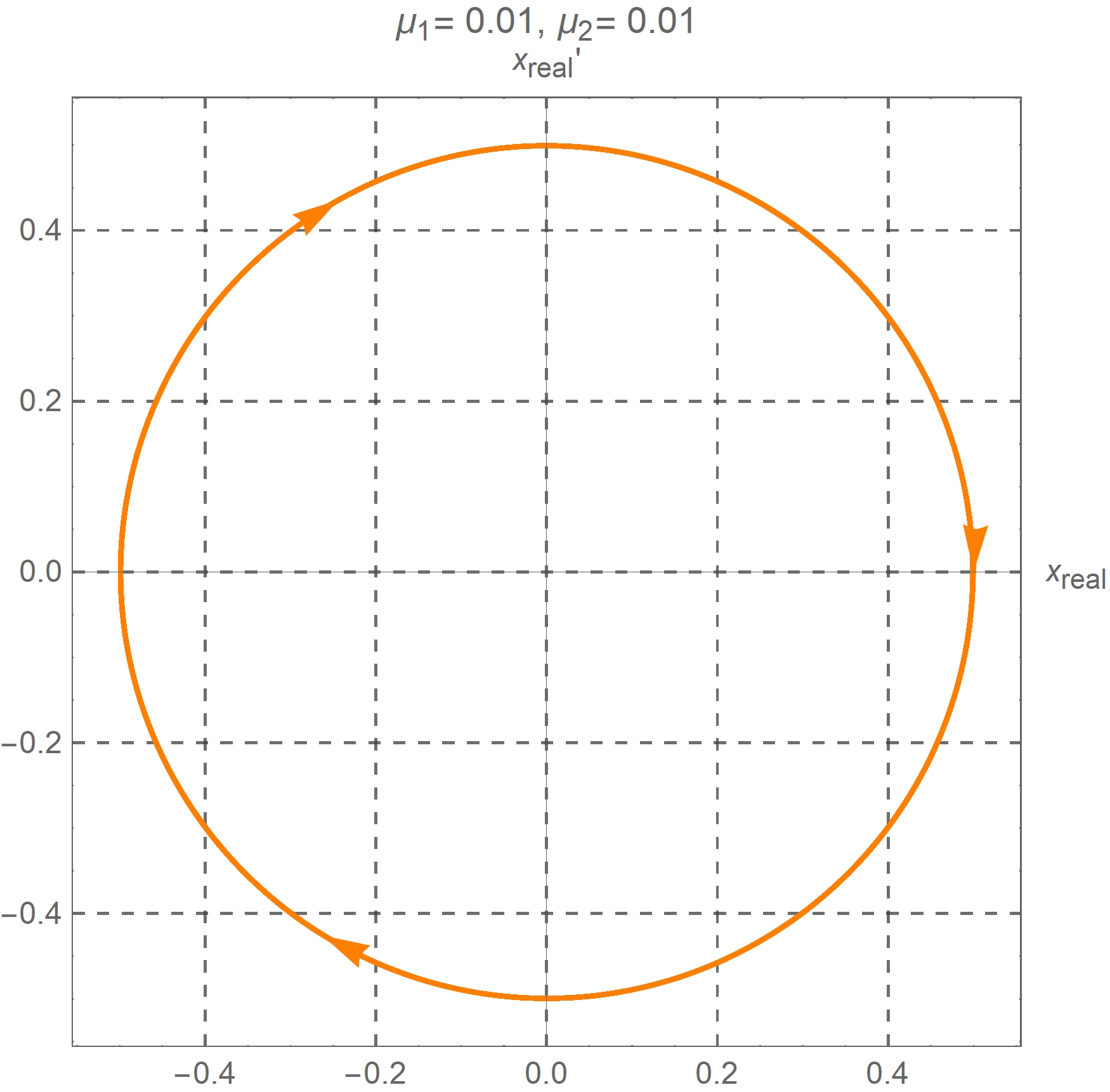}
        }%
        \subfigure[]{%
           \label{fig:second5}
           \includegraphics[width=0.2\textwidth]{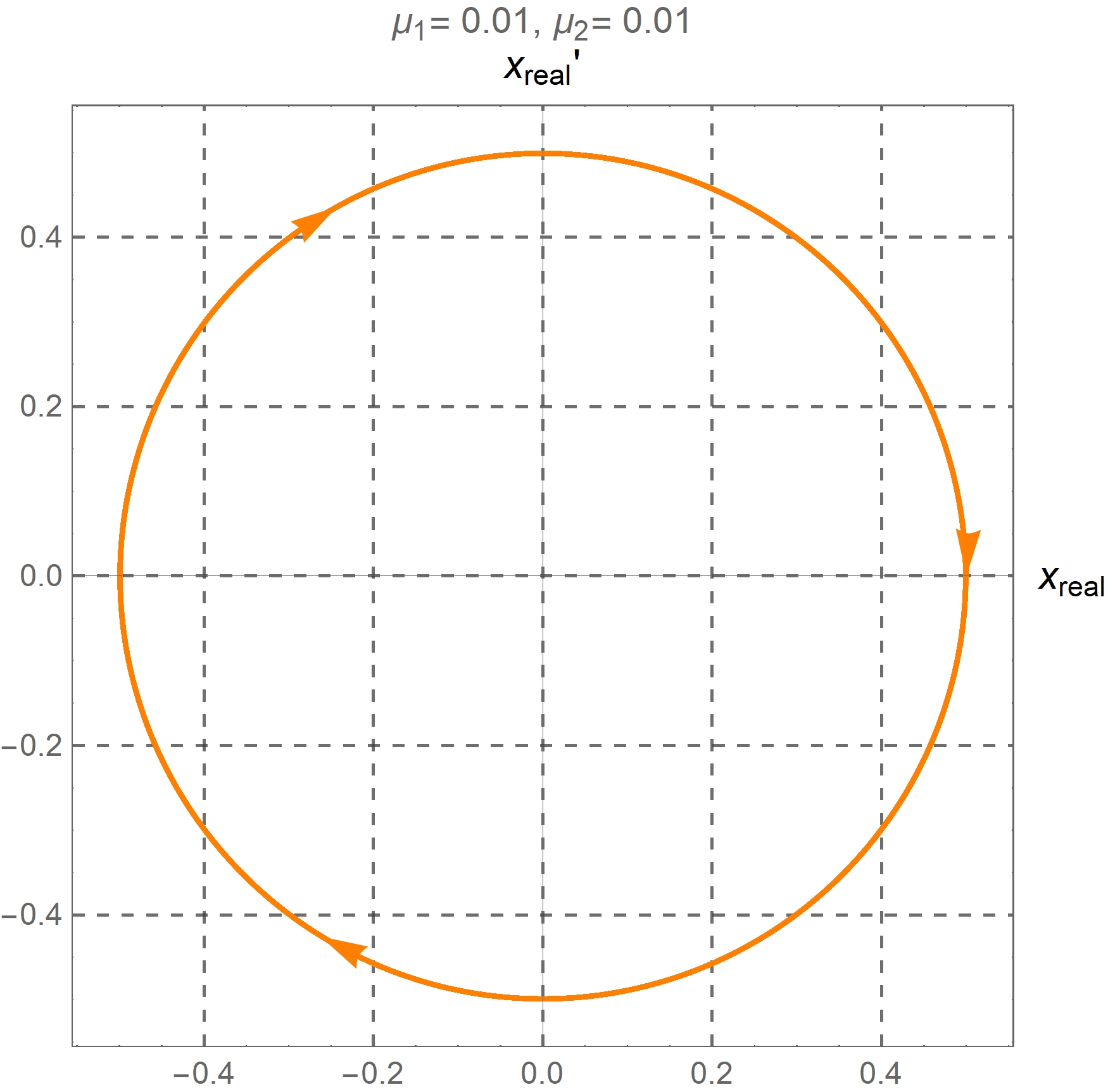}
        } \\
        
    \end{center}
    \caption{Time derivative of real part of $x(t)$ vs real part of $x(t)$ have been plotted for the 
the $\mathcal{PT}$ symmetric case. (a) RG-aided Solution and (b) Numerical Solution. Here $\mu_1=0.01$ and $\mu_2=0.01$ which is $\mu_1=\mu_2 $ case.}
   \label{fig:subfigures i5}
\end{figure}

In this section, we compare the  RG-aided and numerical solutions. We find that the RG-aided solution matches with our numerical solution as is shown in \cref{fig:subfigures i1,fig:subfigures i5}. 

We observe a similar match for the imaginary part of $x(t)$ and also for both real and imaginary parts of $y(t)$. Also in this case we notice limit cycles in both of our solutions.

Now in the $\mathcal{PT}$ symmetric case, we see that the imaginary part of RG aided $x(t)$ and $y(t)$ vanishes. The real part of $y(t)$ follows a similar trend as the real part of $x(t)$. Numerical analysis gives us exactly the same results. Thus we obtain a solid correspondence between the RG-aided solutions and numerical solutions both in $\mathcal{PT}$ symmetric case and non-$\mathcal{PT}$ symmetric case. Also for the $\mathcal{PT}$ symmetric case, we observe that centre solutions exist as seen in \cref{fig:subfigures i5}. This result could again be inferred from \cref{Ataux2} where we see that for the non-$\mathcal{PT}$ symmetric case ($\mu_1\neq\mu_2$) we should find limit cycles and for $\mathcal{PT}$ symmetric case ($\mu_1=\mu_2$) we should get centres as $\dv{A(\tau)}{\tau}$ becomes zero. 

\section{Quantum Mechanical Analysis}\label{QM_MvdP}

For the quantum analysis, we consider the Weyl-ordered version of the classical Hamiltonian given in \cref{Hamiltonian} as,

\begin{align}
    \mathcal{H}&=p_xp_y+\omega^2xy-\iimg\left[\frac{\mu_1}{2}\left(1-x^2\right)\left(yp_y+p_yy\right)\right.\nonumber\\
&\mathrel{\phantom{=}}\left.+\frac{\mu_2}{2}\left(1-y^2\right)\left(xp_x+p_xx\right)\right]\label{Hamiltonian_QM}
\end{align}

\begin{figure}[ht]
     \begin{center}

            \includegraphics[width=0.48\textwidth]{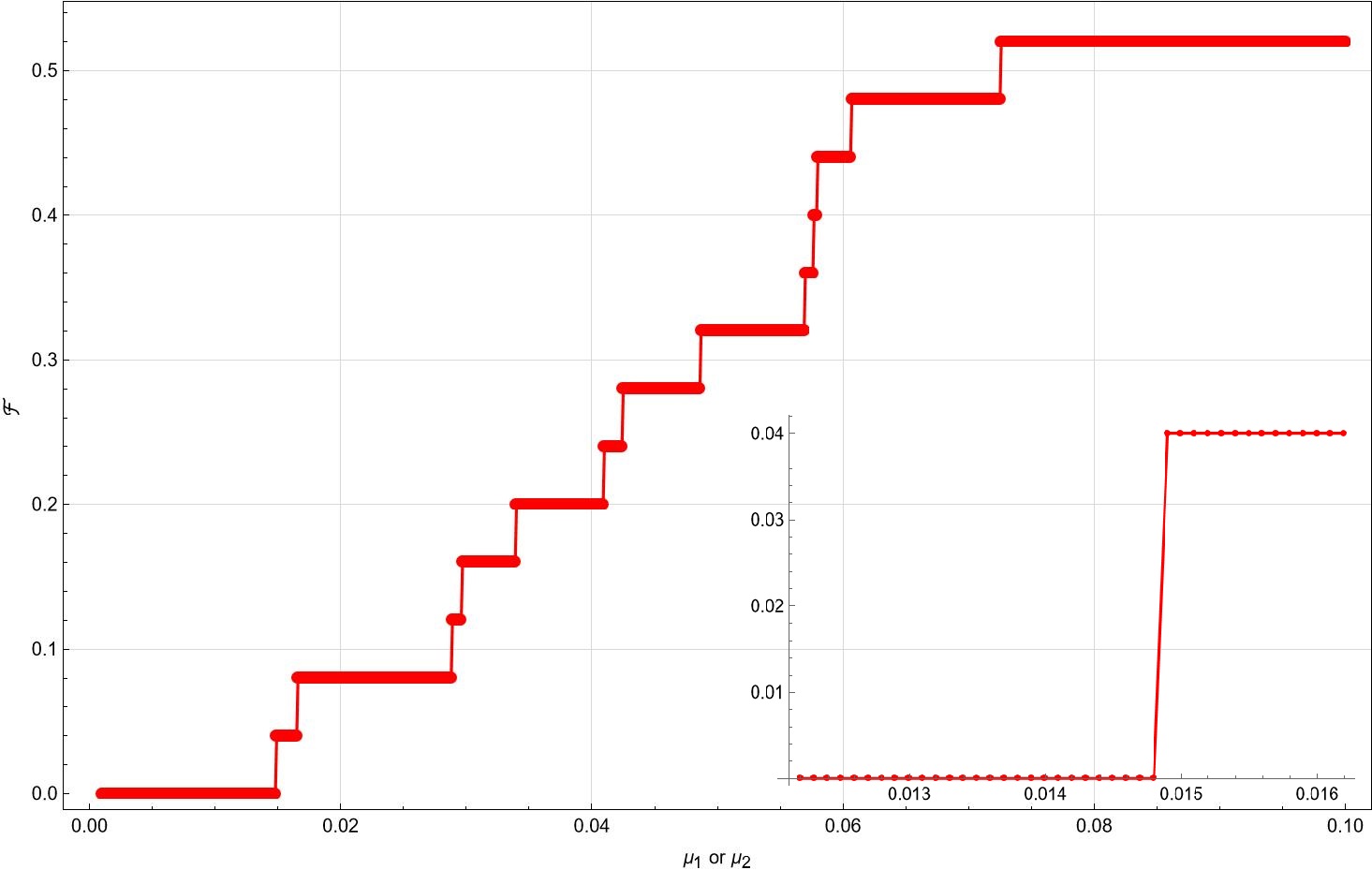}
        
    \end{center}
    \caption{ Fraction of complex values eigenvalues $\mathcal{F}$ vs $\mu$. The phase transition is further illustrated in the inset.}
   \label{fig_PT_transition1}
\end{figure}

We have already mentioned in section \ref{MvdP-section} that this Hamiltonian is non-Hermitian but $\mathcal{PT}$-symmetric when $\mu_1=\mu_2=\mu$. We numerically solve the time-independent Schr\"odinger equation, $\mathcal{H}\psi=E\psi$ for this system to obtain the eigenvalues $(E)$. To demonstrate the $\mathcal{PT}$ phase transition we define a new parameter $\mathcal{F}$ which represents the fraction of complex eigenvalues to the total number of eigenvalues. We have numerically calculated $\mathcal{F}$ and have plotted it with respect to the Hermiticity breaking parameter ($\mu_1 $ or  $\mu_2$) in the fig \ref{fig_PT_transition1} for the $\mathcal{PT}$-symmetric case  ($\mu_1=\mu_2$). We observed that $\mathcal{F}$ remains at zero until we cross a critical ($\mu_c$) value of the Hermiticity breaking parameter. This indicates all eigenvalues are real for $\mu_c\le \mu$. The system passes from the unbroken $\mathcal{PT}$ phase to the broken $\mathcal{PT}$ phase at $\mu=\mu_c$. The eigenvalues occur in complex pairs for $\mu>\mu_c $ that is in the $\mathcal{PT}$ broken region. For the confirmation of $\mathcal{PT}$ phase transition, one needs to show that eigenstates also respect $\mathcal{PT}$ symmetry in the unbroken phase.

We further consider the case $\mu_1\ne \mu_2$ and study the variation of $\mathcal{F}$ with respect to the ration $\frac{\mu_1}{\mu_2}$. We observe that $\mathcal{F}$ becomes zero only when $\frac{\mu_1}{\mu_2}=1$. We consider sufficiently low values of $\mu_1,\, \mu_2$ to be in the unbroken phase when $\mu_1=\mu_2$.

\begin{figure}[ht]
     \begin{center}
            \includegraphics[width=0.48\textwidth]{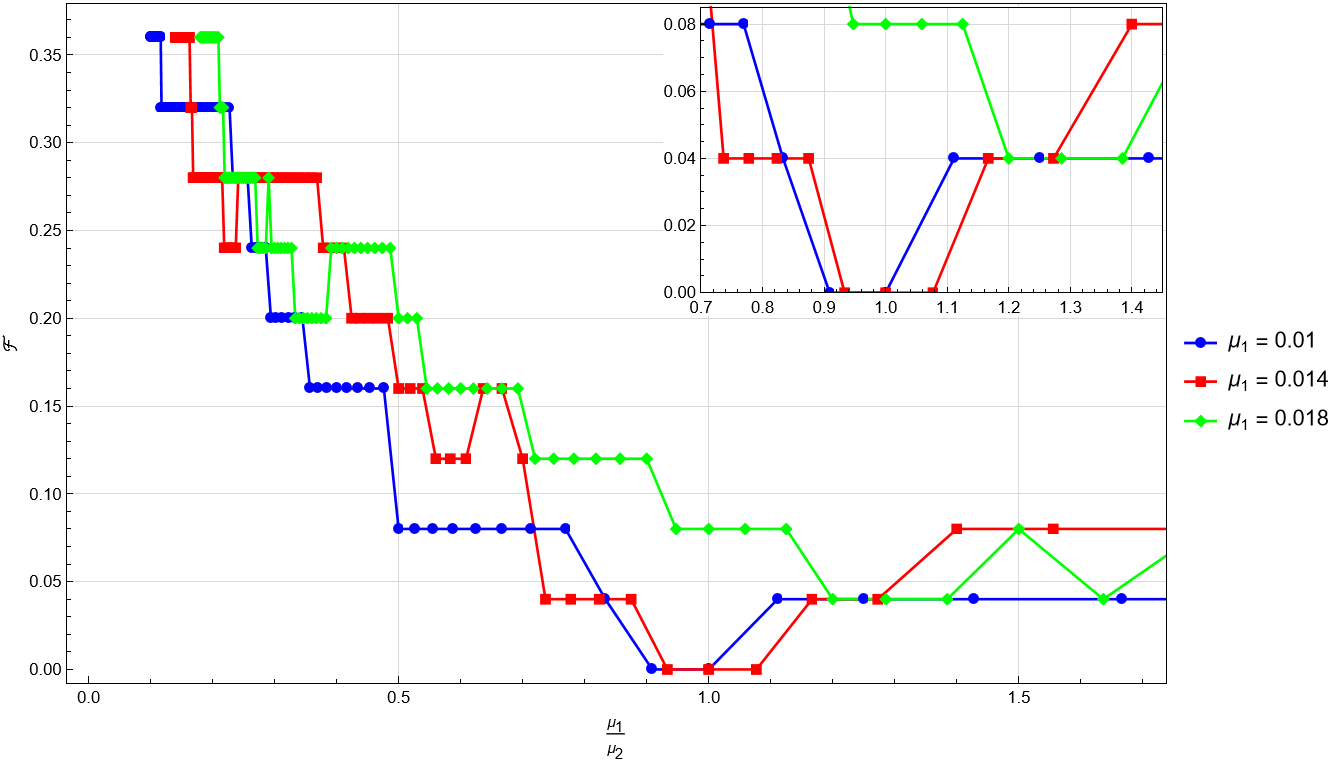}
        
    \end{center}
    \caption{ Fraction of complex values eigenvalues  $\left(\mathcal{F}\right)$ vs $\frac{\mu_1}{\mu_2}$.  $\mu_1$ is kept fixed  and $\mu_2$ is varied. The phase transition is further highlighted in the inset figure.}
   \label{fig_PT_transition2}
\end{figure}

It is also evident from \cref{fig_PT_transition2} when $\mu_1=\mu_2$, $\mathcal{F}$ =0 and the system is in unbroken $\mathcal{PT}$-symmetry phase when $\mu_1$ and $\mu_2$ are sufficiently small. We find the broken-$\mathcal{PT}$ phase in our system when the value of the coupling is greater than the critical value. In that case, even when $\mu_1=\mu_2$, $\mathcal{F}$ has a non-zero value. Hence the 2d non-Hermitian vdP oscillator exhibits $\mathcal{PT}$ phase transition.

\section{Conclusion}\label{conclusion}

In this paper, we have considered a non-Hermitian $\mathcal{PT}$-symmetric modification of the well-known 2d vdP oscillator and studied the dynamics classically as well as in the quantum domain. We have demonstrated that the usual perturbation method fails to give physically acceptable solutions and thus we had to incorporate the renormalization group techniques to get the same. We analytically have obtained the classical solution using RG-technique which is physically acceptable and matches with numerically obtained solutions.  Our classical solutions can be categorized into limit cycles and centres depending on the amplitude of the flow equations. We have limit cycle solutions when the corresponding quantum Hamiltonian is not $\mathcal{PT}$ symmetric and we have stable centre type solutions with fixed amplitude when the corresponding quantum Hamiltonian is $\mathcal{PT}$ symmetric.  We further analyse the system in the quantum domain numerically to study the $\mathcal{PT}$ phase transition. $\mathcal{PT}$-symmetric non-Hermitian 2d vdP oscillator is shown to undergo a $\mathcal{PT}$ phase transition when the Hermiticity breaking parameter exceeds a critical value.

{\bf Acknowledgement:} One of us (BPM) acknowledges the research grant for faculty under IoE Scheme (Number 6031) of Banaras Hindu University, Varanasi.

\appendix
\section{Detailed Solutions}\label{Appendix}

In this appendix, we provide more details on the calculations performed in the paper.

\subsection{Perturbative Solutions}

We considered a perturbative expansion of $x(t)$ and $y(t)$ in terms of $\mu_1$ and $\mu_2$ as shown in \cref{expansion_pert} to find the solution for \cref{xddot,yddot} order by order in $\mu_1$ and $\mu_2$. For $\mathcal{O}\left(\mu_1^0\mu_2^0\right)$ we have seen equations \cref{ddotx00,y_00}. Using the initial conditions as given in section \ref{PertSoln}, we find using Mathematica the solutions of \cref{ddotx00,y_00} as given by,

\begin{align}
	x_{00}\left(t\right)&=A(t_0) \cos(\omega(t-t_0))\label{x_00_sol}\\
	y_{00}\left(t\right)&=B(t_0) \cos(\omega(t-t_0))\label{y_00_sol}
\end{align}

Substituting these solutions from \cref{x_00_sol,y_00_sol} in \cref{ddotx10,ddotx01} and \cref{y_10,y_01} then using the initial conditions which are explicitly given in section \ref{PertSoln}, we find the solutions to $x_{10}\left(t\right)$, $x_{01}\left(t\right)$ and $y_{10}\left(t\right)$, $y_{01}\left(t\right)$ using Mathematica to be of the following form,
\begin{align}
	x_{10}(t)&=-\frac{\iimg A(t_0)}{8\omega}\left[2(-4+A^2(t_0))(t-t_0)\right.\nonumber \\
	&\mathrel{\phantom{=}}\left.\times\omega\cos(\omega(t-t_0))\right.\nonumber \\
	&\mathrel{\phantom{=}}\left.+(8-3A^2(t_0)+A^2(t_0)\cos(2\omega(t-t_0)))\right.\nonumber \\
	&\mathrel{\phantom{=}}\left.\times\sin(\omega(t-t_0))\right]\label{ddotx10s} \\
	x_{01}(t)&=\frac{\iimg A(t_0)}{8\omega}\left[2(-4+B^2(t_0))(t-t_0)\omega\cos(\omega(t-t_0))\right.\nonumber \\
	&\mathrel{\phantom{=}} \left.+(8-3B^2(t_0)+B^2(t_0)\cos(2\omega(t-t_0)))\right.\nonumber \\
	&\mathrel{\phantom{=}}\left.\times\sin(\omega(t-t_0))\right]\label{ddotx01s}\\
	y_{10}(t)&=\frac{\iimg B(t_0)}{8\omega} \left[2\left(-4+A^2(t_0)\right)(t-t_0)\omega\cos(\omega(t-t_0))\right.\nonumber\\
	&\mathrel{\phantom{=}}	\left.+\left(8-3A^2(t_0)+A^2(t_0)\cos\left(2\omega(t-t_0)\right)\right)\right.\nonumber \\
	&\mathrel{\phantom{=}}\left.\times\sin\left(\omega(t-t_0)\right)\right]\\
	y_{01}(t)&=\frac{-\iimg B(t_0)}{8\omega} \left[2(-4+B^2(t_0))(t-t_0)\right.\nonumber \\
	&\mathrel{\phantom{=}}\left.\times\omega\cos(\omega(t-t_0))\right.\nonumber\\
	&\mathrel{\phantom{=}}\left.+\left(8-3B^2(t_0)+B^2(t_0)\cos\left(2\omega(t-t_0)\right)\right)\right.\nonumber \\
	&\mathrel{\phantom{=}}\left.\times\sin\left(\omega(t-t_0)\right)\right]
\end{align}

We use the above results to write down the complete perturbative solution for $x(t)$ and $y(t)$ up to $\mathcal{O}\left(\mu_1\right)$ and $\mathcal{O}\left(\mu_2\right)$ in \cref{x_sol,y_sol}.

\subsection{Flow Equation's Details}

After finding the RG-aided solutions for $x(t)$ as shown in \cref{x_sol_RG}, we use them in our flow equation given by \cref{deri_x_0} to find the coupled DEs followed by $A\left(\tau\right)$, $B\left(\tau\right)$ and $\theta\left(\tau\right)$. This we do by equating the coefficients of $\cos(\omega(t+\theta(\tau)))$, $\sin(\omega(t+\theta(\tau)))$, $\cos(3\omega(t+\theta(\tau)))$ and $\sin(3\omega(t+\theta(\tau)))$ separately to zero. And thus giving us the following coupled DEs of the following form,
\begin{align}
\dv{A(\tau)}{\tau}+\dv{\theta(\tau)}{\tau}\left[\frac{-\iimg A(\tau)}{2}(\mu_1-\mu_2)+\frac{7\iimg\mu_1A^3(\tau)}{32}\right. \nonumber \\
\left.-\frac{7\iimg\mu_2A(\tau)B^2(\tau)}{32}\right]+\frac{\iimg}{8}A(\tau)\left[\mu_1(-4+A^2(\tau))\right.\nonumber\\
\left.-\mu_2(-4+B^2(\tau))\right]&=0\label{fe_1} \\ 
\dv{A(\tau)}{\tau}\left[-\frac{\iimg}{2\omega^2}(\mu_1-\mu_2)+\frac{21\iimg\mu_1}{32\omega^2}A^2(\tau)\right. \nonumber \\
\left.-\frac{7\iimg\mu_2}{32\omega^2}B^2(\tau) \right]-\dv{B(\tau)}{\tau}\left[\frac{7\iimg\mu_2}{16\omega^2}A(\tau)B(\tau) \right]\nonumber \\
-A(\tau)\dv{\theta(\tau)}{\tau}&=0\label{fe_2}\\
\dv{A(\tau)}{\tau}\left[3\mu_1A^2(\tau)-\mu_2B^2(\tau) \right]\nonumber\\
-\dv{B(\tau)}{\tau}\left[2\mu_2A(\tau)B(\tau) \right]&=0\label{fe_3}\\
\dv{\theta(\tau)}{\tau}\left[-\mu_2A(\tau)B^2(\tau)+\mu_1A^3(\tau) \right]&=0\label{fe_4}
\end{align}
To solve these coupled ODEs we consider the \cref{fe_4} and find that we can decompose it into the following two conditions,

\begin{align}
\dv{\theta(\tau)}{\tau}=0 \label{x_case1}\\
B(\tau)=\mp \sqrt{\frac{\mu_1}{\mu_2}}A(\tau)\label{x_case2}
\end{align}

We then use two conditions separately to find two separate solutions for these coupled ODEs. So firstly we consider, \\
Case 1: $\dv{\theta(\tau)}{\tau}=0$\\
We use this given condition to find that \cref{fe_1,fe_2,fe_3} could be reduced to the following uncoupled ODEs,

\begin{align}
\dv{\theta(\tau)}{\tau}=0\label{Ataux1}\\
\dv{A(\tau)}{\tau}=0\label{Btaux1}\\
\dv{B(\tau)}{\tau}=0\label{Ttaux1}
\end{align}
Now to solve the above equations, we consider the following initial conditions,
\begin{align}
\theta(\tau=t_0)&=-t_0\\
A(\tau=t_0)&=A(t_0)=\text{constant}\\
B(\tau=t_0)&=B(t_0)=\text{constant}
\end{align}

So trivially we find in accordance with the above initial conditions the solutions for \cref{Ataux1,Btaux1,Ttaux1} to be of the following form,

\begin{align}
\theta(\tau)&=-t_0\\
A(\tau)&=A(t_0)\\
B(\tau)&=B(t_0)
\end{align}

Thus we see that the total solution of $x(t)$ could be written down in the following form,
\begin{align}
x(t)&=A(t_0)\cos(\omega(t-t_0)) \nonumber \\
&\mathrel{\phantom{=}} -\frac{\iimg\mu_1}{16}A(t_0)\left[2(-4+A^2(t_0))(t-\tau)\right]\cos(\omega(t-t_0))\nonumber \\
&\mathrel{\phantom{=}} -\frac{\iimg\mu_1}{16\omega}A(t_0)\left[8-3A^2(t_0)+A^2(t_0)\right]\nonumber \\
&\mathrel{\phantom{=}}\times\cos(2\omega(t-t_0))\sin(\omega(t-t_0))\nonumber \\
&\mathrel{\phantom{=}} +\frac{\iimg\mu_2}{16}A(t_0)\left[2(-4+B^2(t_0))(t-\tau)\right]\cos(\omega(t-t_0))\nonumber \\
&\mathrel{\phantom{=}} +\frac{\iimg\mu_2}{16\omega}A(t_0)\left[8-3B^2(t_0)+B^2(t_0)\right]\nonumber \\
&\mathrel{\phantom{=}}\times\cos(2\omega(t-t_0))\sin(\omega(t-t_0))
\end{align}
As $\tau$ is considered arbitrarily, the most convenient choice would be to consider it to be equal to $t$ such that the above equation takes the form as given in \cref{x_sol_RGfinal_1} which is the complete RG-aided solution for $x(t)$ for this given case. Now we consider the other condition in the following way,\\
Case 2: $B(\tau)=\mp \sqrt{\frac{\mu_1}{\mu_2}}A(\tau)$\\
We use this condition similarly as before to simplify \cref{fe_1,fe_2,fe_3} to the following two ODEs,
\begin{align}
\dv{A(\tau)}{\tau}\left[1-\frac{(\mu_1-\mu_2)^2}{4\omega^2} \right]&=\frac{\iimg(\mu_1-\mu_2)}{2}A(\tau)\label{Ataux2}\\
\dv{\theta(\tau)}{\tau}&=\frac{(\mu_1-\mu_2)^2}{4\omega^2-(\mu_1-\mu_2)^2}\label{Ttaux2}\tau
\end{align}
Now to solve the above ODE we choose the following initial condition,
\begin{align}
\theta(\tau=t_0)&=-t_0\\
A(\tau=t_0)&=A(t_0)=\text{constant}\\
B(\tau=t_0)&=B(t_0)=\text{constant}
\end{align}
Thus using these initial conditions we find the solution to \cref{Ataux1,Ttaux2} as the following,
\begin{align}
A(\tau)&=A(t_0)\exp(\frac{2\iimg(\mu_1-\mu_2)\omega^2}{4\omega^2-(\mu_1-\mu_2)^2}(\tau-t_0))\\
B(\tau)&=\mp\sqrt{\frac{\mu_1}{\mu_2}} A(t_0)\exp(\frac{2\iimg(\mu_1-\mu_2)\omega^2}{4\omega^2-(\mu_1-\mu_2)^2}(\tau-t_0))\\
\theta(\tau)&=\frac{(\mu_1-\mu_2)^2\tau-4\omega^2t_0}{4\omega^2-(\mu_1-\mu_2)^2}
\end{align}

Note that to find the solution for $B\left(\tau\right)$ we considered the fact that for this case $A\left(\tau\right)$ is related to $B\left(\tau\right)$. We now use these above results to write the final solution for $x(t)$ but again using the fact that $\tau$ can be chosen arbitrarily and for convenience we choose it to be equal to $t$. Thus giving us the full RG-aided solution for $x(t)$ as shown in \cref{x_sol_RGfinal_2}. 

Again to find the full RG-aided solution of $y(t)$ we perform an exactly similar calculation thus giving us two solutions as shown by \cref{y_sol_RGfinal_1,y_sol_RGfinal_2}.

\bibliographystyle{unsrtnat}
\bibliography{citation.bib}

\begin{thebibliography}{52}
\providecommand{\natexlab}[1]{#1}
\providecommand{\url}[1]{\texttt{#1}}
\expandafter\ifx\csname urlstyle\endcsname\relax
  \providecommand{\doi}[1]{doi: #1}\else
  \providecommand{\doi}{doi: \begingroup \urlstyle{rm}\Url}\fi

\bibitem[Van~der Pol(1926)]{van1926lxxxviii}
Balth Van~der Pol.
\newblock Lxxxviii. on “relaxation-oscillations”.
\newblock \emph{The London, Edinburgh, and Dublin Philosophical Magazine and
  Journal of Science}, 2\penalty0 (11):\penalty0 978--992, 1926.

\bibitem[Robinson(1987)]{robinson1987modified}
FNH Robinson.
\newblock The modified van der pol oscillator.
\newblock \emph{IMA journal of applied mathematics}, 38\penalty0 (2):\penalty0
  135--150, 1987.

\bibitem[Cartwright et~al.(1999)Cartwright, Egu{\'\i}luz,
  Hern{\'a}ndez-Garc{\'\i}a, and Piro]{cartwright1999dynamics}
Julyan~HE Cartwright, V{\'\i}ctor~M Egu{\'\i}luz, Emilio
  Hern{\'a}ndez-Garc{\'\i}a, and Oreste Piro.
\newblock Dynamics of elastic excitable media.
\newblock \emph{International Journal of Bifurcation and Chaos}, 9\penalty0
  (11):\penalty0 2197--2202, 1999.

\bibitem[Guckenheimer et~al.(2000)Guckenheimer, Hoffman, and
  Weckesser]{guckenheimer2000numerical}
John Guckenheimer, Kathleen Hoffman, and Warren Weckesser.
\newblock Numerical computation of canards.
\newblock \emph{International Journal of Bifurcation and Chaos}, 10\penalty0
  (12):\penalty0 2669--2687, 2000.

\bibitem[Bender and Boettcher(1998)]{bender1998real}
Carl~M Bender and Stefan Boettcher.
\newblock Real spectra in non-hermitian hamiltonians having pt symmetry.
\newblock \emph{Physical review letters}, 80\penalty0 (24):\penalty0 5243,
  1998.

\bibitem[Bender(2007)]{bender2007making}
Carl~M Bender.
\newblock Making sense of non-hermitian hamiltonians.
\newblock \emph{Reports on Progress in Physics}, 70\penalty0 (6):\penalty0 947,
  2007.

\bibitem[Mostafazadeh(2010)]{mostafazadeh2010pseudo}
Ali Mostafazadeh.
\newblock Pseudo-hermitian representation of quantum mechanics.
\newblock \emph{International Journal of Geometric Methods in Modern Physics},
  7\penalty0 (07):\penalty0 1191--1306, 2010.

\bibitem[Khare and Mandal(2000)]{khare2000pt}
Avinash Khare and Bhabani~Prasad Mandal.
\newblock A pt-invariant potential with complex qes eigenvalues.
\newblock \emph{Physics Letters A}, 272\penalty0 (1-2):\penalty0 53--56, 2000.

\bibitem[Bender et~al.(2002)Bender, Brody, and Jones]{bender2002complex}
Carl~M Bender, Dorje~C Brody, and Hugh~F Jones.
\newblock Complex extension of quantum mechanics.
\newblock \emph{Physical Review Letters}, 89\penalty0 (27):\penalty0 270401,
  2002.

\bibitem[Das and Greenwood(2010)]{das2010alternative}
Ashok Das and L~Greenwood.
\newblock An alternative construction of the positive inner product for
  pseudo-hermitian hamiltonians: Examples.
\newblock \emph{Journal of Mathematical Physics}, 51\penalty0 (4), 2010.

\bibitem[Mostafazadeh(2007)]{mostafazadeh2007time}
Ali Mostafazadeh.
\newblock Time-dependent pseudo-hermitian hamiltonians defining a unitary
  quantum system and uniqueness of the metric operator.
\newblock \emph{Physics Letters B}, 650\penalty0 (2-3):\penalty0 208--212,
  2007.

\bibitem[Ju et~al.(2019)Ju, Miranowicz, Chen, and Nori]{ju2019non}
Chia-Yi Ju, Adam Miranowicz, Guang-Yin Chen, and Franco Nori.
\newblock Non-hermitian hamiltonians and no-go theorems in quantum information.
\newblock \emph{Physical Review A}, 100\penalty0 (6):\penalty0 062118, 2019.

\bibitem[Gardas et~al.(2016)Gardas, Deffner, and Saxena]{gardas2016non}
Bart{\l}omiej Gardas, Sebastian Deffner, and Avadh Saxena.
\newblock Non-hermitian quantum thermodynamics.
\newblock \emph{Scientific reports}, 6\penalty0 (1):\penalty0 23408, 2016.

\bibitem[Tzeng et~al.(2021)Tzeng, Ju, Chen, and Huang]{tzeng2021hunting}
Yu-Chin Tzeng, Chia-Yi Ju, Guang-Yin Chen, and Wen-Min Huang.
\newblock Hunting for the non-hermitian exceptional points with fidelity
  susceptibility.
\newblock \emph{Physical Review Research}, 3\penalty0 (1):\penalty0 013015,
  2021.

\bibitem[Bender et~al.(1999{\natexlab{a}})Bender, Boettcher, Jones, and
  Savage]{Bender_1999}
Carl~M Bender, Stefan Boettcher, H~F Jones, and Van~M Savage.
\newblock Complex square well - a new exactly solvable quantum mechanical
  model.
\newblock \emph{Journal of Physics A: Mathematical and General}, 32\penalty0
  (39):\penalty0 6771--6781, sep 1999{\natexlab{a}}.
\newblock \doi{10.1088/0305-4470/32/39/305}.
\newblock URL \url{https://doi.org/10.1088%2F0305-4470%2F32%2F39%2F305}.

\bibitem[Bender et~al.(1999{\natexlab{b}})Bender, Boettcher, and
  Meisinger]{BENDER_PTSymmetricQM}
Carl~M. Bender, Stefan Boettcher, and Peter~N. Meisinger.
\newblock {$\mathcal{PT}$-symmetric quantum mechanics}.
\newblock \emph{Journal of Mathematical Physics}, 40\penalty0 (5):\penalty0
  2201--2229, 05 1999{\natexlab{b}}.
\newblock ISSN 0022-2488.
\newblock \doi{10.1063/1.532860}.
\newblock URL \url{https://doi.org/10.1063/1.532860}.

\bibitem[Ghatak et~al.(2013)Ghatak, Mandal, and Mandal]{ghatak2013various}
Ananya Ghatak, Raka Dona~Ray Mandal, and Bhabani~Prasad Mandal.
\newblock Various scattering properties of a new pt-symmetric non-hermitian
  potential.
\newblock \emph{Annals of Physics}, 336:\penalty0 540--552, 2013.

\bibitem[Modak and Mandal(2021)]{modak2021eigenstate}
Ranjan Modak and Bhabani~Prasad Mandal.
\newblock Eigenstate entanglement entropy in a pt-invariant non-hermitian
  system.
\newblock \emph{Physical Review A}, 103\penalty0 (6):\penalty0 062416, 2021.

\bibitem[Hasan et~al.(2020)Hasan, Singh, and Mandal]{hasan2020role}
Mohammad Hasan, Vibhav~Narayan Singh, and Bhabani~Prasad Mandal.
\newblock Role of pt-symmetry in understanding hartman effect.
\newblock \emph{The European Physical Journal Plus}, 135:\penalty0 1--25, 2020.

\bibitem[Shukla et~al.(2023)Shukla, Modak, and Mandal]{shukla2023uncertainty}
Namrata Shukla, Ranjan Modak, and Bhabani~Prasad Mandal.
\newblock Uncertainty relation for non-hermitian systems.
\newblock \emph{Physical Review A}, 107\penalty0 (4):\penalty0 042201, 2023.

\bibitem[Khare and Mandal(2009)]{khare2009new}
Avinash Khare and Bhabani~Prasad Mandal.
\newblock New quasi-exactly solvable hermitian as well as
  non-hermitian-invariant potentials.
\newblock \emph{Pramana}, 73\penalty0 (2):\penalty0 387--395, 2009.

\bibitem[Yadav et~al.(2016)Yadav, Khare, Bagchi, Kumari, and
  Mandal]{yadav2016parametric}
Rajesh~Kumar Yadav, Avinash Khare, Bijan Bagchi, Nisha Kumari, and
  Bhabani~Prasad Mandal.
\newblock Parametric symmetries in exactly solvable real and pt-symmetric
  complex potentials.
\newblock \emph{Journal of Mathematical Physics}, 57\penalty0 (6), 2016.

\bibitem[Basu-Mallick et~al.(2005)Basu-Mallick, Bhattacharyya, and
  Mandal]{basu2005phase}
B~Basu-Mallick, Tanaya Bhattacharyya, and Bhabani~Prasad Mandal.
\newblock Phase shift analysis of pt-symmetric non-hermitian extension of an-1
  calogero model without confining interaction.
\newblock \emph{Modern Physics Letters A}, 20\penalty0 (07):\penalty0 543--552,
  2005.

\bibitem[Ohlsson and Zhou(2020)]{ohlsson2020transition}
Tommy Ohlsson and Shun Zhou.
\newblock Transition probabilities in the two-level quantum system with
  pt-symmetric non-hermitian hamiltonians.
\newblock \emph{Journal of Mathematical Physics}, 61\penalty0 (5), 2020.

\bibitem[Klauck et~al.(2019)Klauck, Teuber, Ornigotti, Heinrich, Scheel, and
  Szameit]{klauck2019observation}
F~Klauck, Lucas Teuber, Marco Ornigotti, Matthias Heinrich, Stefan Scheel, and
  Alexander Szameit.
\newblock Observation of pt-symmetric quantum interference.
\newblock \emph{Nature Photonics}, 13\penalty0 (12):\penalty0 883--887, 2019.

\bibitem[Ashida et~al.(2017)Ashida, Furukawa, and Ueda]{ashida2017parity}
Yuto Ashida, Shunsuke Furukawa, and Masahito Ueda.
\newblock Parity-time-symmetric quantum critical phenomena.
\newblock \emph{Nature communications}, 8\penalty0 (1):\penalty0 15791, 2017.

\bibitem[Zhang et~al.(2019)Zhang, Wang, and Gong]{zhang2019time}
Da-Jian Zhang, Qing-hai Wang, and Jiangbin Gong.
\newblock Time-dependent pt-symmetric quantum mechanics in generic
  non-hermitian systems.
\newblock \emph{Physical Review A}, 100\penalty0 (6):\penalty0 062121, 2019.

\bibitem[Bagarello et~al.(2018)Bagarello, Gargano, and
  Spagnolo]{bagarello2018bi}
Fabio Bagarello, Francesco Gargano, and Salvatore Spagnolo.
\newblock Bi-squeezed states arising from pseudo-bosons.
\newblock \emph{Journal of Physics A: Mathematical and Theoretical},
  51\penalty0 (45):\penalty0 455204, 2018.

\bibitem[Bagarello and Hatano(2021)]{bagarello2021chain}
Fabio Bagarello and Naomichi Hatano.
\newblock A chain of solvable non-hermitian hamiltonians constructed by a
  series of metric operators.
\newblock \emph{Annals of Physics}, 430:\penalty0 168511, 2021.

\bibitem[Pal et~al.(2022)Pal, Modak, and Mandal]{pal2022dna}
Tanmoy Pal, Ranjan Modak, and Bhabani~Prasad Mandal.
\newblock Dna unzipping as $\{$$\backslash$bf PT$\}$-symmetry breaking
  transition.
\newblock \emph{arXiv preprint arXiv:2212.14394}, 2022.

\bibitem[Klaiman et~al.(2008)Klaiman, G{\"u}nther, and
  Moiseyev]{klaiman2008visualization}
Shachar Klaiman, Uwe G{\"u}nther, and Nimrod Moiseyev.
\newblock Visualization of branch points in pt-symmetric waveguides.
\newblock \emph{Physical review letters}, 101\penalty0 (8):\penalty0 080402,
  2008.

\bibitem[Mandal et~al.(2015)Mandal, Mourya, Ali, and Ghatak]{mandal2015pt}
Bhabani~Prasad Mandal, Brijesh~K Mourya, Kawsar Ali, and Ananya Ghatak.
\newblock Pt phase transition in a (2+ 1)d relativistic system.
\newblock \emph{Annals of Physics}, 363:\penalty0 185--193, 2015.

\bibitem[Xu et~al.(2016)Xu, Mason, Jiang, and Harris]{xu2016topological}
Haitan Xu, David Mason, Luyao Jiang, and JGE Harris.
\newblock Topological energy transfer in an optomechanical system with
  exceptional points.
\newblock \emph{Nature}, 537\penalty0 (7618):\penalty0 80--83, 2016.

\bibitem[Xiao et~al.(2016)Xiao, Lin, and Fan]{xiao2016hyperbolic}
Meng Xiao, Qian Lin, and Shanhui Fan.
\newblock Hyperbolic weyl point in reciprocal chiral metamaterials.
\newblock \emph{Physical review letters}, 117\penalty0 (5):\penalty0 057401,
  2016.

\bibitem[Musslimani et~al.(2008)Musslimani, Makris, El-Ganainy, and
  Christodoulides]{musslimani2008optical}
ZH~Musslimani, Konstantinos~G Makris, Ramy El-Ganainy, and Demetrios~N
  Christodoulides.
\newblock Optical solitons in p t periodic potentials.
\newblock \emph{Physical Review Letters}, 100\penalty0 (3):\penalty0 030402,
  2008.

\bibitem[Guo et~al.(2009)Guo, Salamo, Duchesne, Morandotti, Volatier-Ravat,
  Aimez, Siviloglou, and Christodoulides]{guo2009observation}
A~Guo, GJ~Salamo, D~Duchesne, R~Morandotti, M~Volatier-Ravat, V~Aimez,
  GA~Siviloglou, and DN~Christodoulides.
\newblock Observation of pt-symmetry breaking in complex optical potentials.
\newblock \emph{Physical review letters}, 103\penalty0 (9):\penalty0 093902,
  2009.

\bibitem[West et~al.(2010)West, Kottos, and Prosen]{west2010p}
Carl~T West, Tsampikos Kottos, and Toma{\v{z}} Prosen.
\newblock P t-symmetric wave chaos.
\newblock \emph{Physical review letters}, 104\penalty0 (5):\penalty0 054102,
  2010.

\bibitem[Pan et~al.(2018)Pan, Zhao, Miao, Longhi, and Feng]{pan2018photonic}
Mingsen Pan, Han Zhao, Pei Miao, Stefano Longhi, and Liang Feng.
\newblock Photonic zero mode in a non-hermitian photonic lattice.
\newblock \emph{Nature communications}, 9\penalty0 (1):\penalty0 1308, 2018.

\bibitem[{\"O}zdemir et~al.(2019){\"O}zdemir, Rotter, Nori, and
  Yang]{ozdemir2019parity}
{\c{S}}ahin~Kaya {\"O}zdemir, Stefan Rotter, Franco Nori, and L~Yang.
\newblock Parity-time symmetry and exceptional points in photonics.
\newblock \emph{Nature materials}, 18\penalty0 (8):\penalty0 783--798, 2019.

\bibitem[Fleury et~al.(2015)Fleury, Sounas, and Al{\`u}]{fleury2015invisible}
Romain Fleury, Dimitrios Sounas, and Andrea Al{\`u}.
\newblock An invisible acoustic sensor based on parity-time symmetry.
\newblock \emph{Nature communications}, 6\penalty0 (1):\penalty0 5905, 2015.

\bibitem[Raval and Prasad~Mandal(2018)]{raval2018deconfinement}
Haresh Raval and Bhabani Prasad~Mandal.
\newblock Deconfinement to confinement as pt phase transition.
\newblock In \emph{DAE-BRNS High Energy Physics Symposium}, pages 617--630.
  Springer, 2018.

\bibitem[Mandal et~al.(2013)Mandal, Mourya, and Yadav]{mandal2013pt}
Bhabani~Prasad Mandal, Brijesh~Kumar Mourya, and Rajesh~Kumar Yadav.
\newblock Pt phase transition in higher-dimensional quantum systems.
\newblock \emph{Physics Letters A}, 377\penalty0 (14):\penalty0 1043--1046,
  2013.

\bibitem[Wilson(1983)]{wilson1983renormalization}
Kenneth~G Wilson.
\newblock The renormalization group and critical phenomena.
\newblock \emph{Reviews of Modern Physics}, 55\penalty0 (3):\penalty0 583,
  1983.

\bibitem[Pelissetto and Vicari(2002)]{pelissetto2002critical}
Andrea Pelissetto and Ettore Vicari.
\newblock Critical phenomena and renormalization-group theory.
\newblock \emph{Physics Reports}, 368\penalty0 (6):\penalty0 549--727, 2002.

\bibitem[Weinberg(1973)]{weinberg1973new}
Steven Weinberg.
\newblock New approach to the renormalization group.
\newblock \emph{Physical Review D}, 8\penalty0 (10):\penalty0 3497, 1973.

\bibitem[Zinn-Justin(2021)]{zinn2021quantum}
Jean Zinn-Justin.
\newblock \emph{Quantum field theory and critical phenomena}, volume 171.
\newblock Oxford University Press, 2021.

\bibitem[Delamotte(2004)]{delamotte2004hint}
Bertrand Delamotte.
\newblock A hint of renormalization.
\newblock \emph{American Journal of Physics}, 72\penalty0 (2):\penalty0
  170--184, 2004.

\bibitem[Bhattacharjee and Ray(2016)]{10.1119/1.4944701}
J.~K. Bhattacharjee and D.~S. Ray.
\newblock {Time-dependent perturbation theory in quantum mechanics and the
  renormalization group}.
\newblock \emph{American Journal of Physics}, 84\penalty0 (6):\penalty0
  434--442, 06 2016.
\newblock ISSN 0002-9505.
\newblock \doi{10.1119/1.4944701}.
\newblock URL \url{https://doi.org/10.1119/1.4944701}.

\bibitem[Strogatz(2018)]{strogatz2018nonlinear}
Steven~H Strogatz.
\newblock \emph{Nonlinear Dynamics and Chaos with Student Solutions Manual:
  With Applications to Physics, Biology, Chemistry, and Engineering}.
\newblock CRC press, 2018.

\bibitem[He(2005)]{he2005limit}
Ji-Huan He.
\newblock Limit cycle and bifurcation of nonlinear problems.
\newblock \emph{Chaos, Solitons \& Fractals}, 26\penalty0 (3):\penalty0
  827--833, 2005.

\bibitem[Sarkar et~al.(2011)Sarkar, Bhattacharjee, Chakraborty, and
  Banerjee]{Sarkar_2011}
A.~Sarkar, J.~K. Bhattacharjee, S.~Chakraborty, and D.~B. Banerjee.
\newblock Center or limit cycle: renormalization group as a probe.
\newblock \emph{The European Physical Journal D}, 64\penalty0 (2-3):\penalty0
  479--489, aug 2011.
\newblock \doi{10.1140/epjd/e2011-20060-1}.
\newblock URL \url{https://doi.org/10.1140%2Fepjd%2Fe2011-20060-1}.

\bibitem[Sarkar and Bhattacharjee(2012)]{Sarkar_2012}
A.~Sarkar and J.K. Bhattacharjee.
\newblock Renormalisation group and isochronous oscillations.
\newblock \emph{The European Physical Journal D}, 66\penalty0 (6), jun 2012.
\newblock \doi{10.1140/epjd/e2012-20427-8}.
\newblock URL \url{https://doi.org/10.1140%2Fepjd%2Fe2012-20427-8}.

\end{thebibliography}
\end{document}